\documentclass[a4paper]{article}
\usepackage{amsmath,graphicx,hyperref,typeface,float,enumitem,multirow,booktabs,array,amssymb,tabularx}
\usepackage{caption}
\usepackage{subcaption}
\usepackage{INTERSPEECH2022}

\title{A Multi-Scale Time-Frequency Spectrogram Discriminator for \\ GAN-based Non-Autoregressive TTS}
\name{Haohan Guo, Hui Lu, Xixin Wu, Helen Meng}
\address{The Chinese University of Hong Kong, Hong Kong SAR, China}
\email{
  \href{mailto:hguo@se.cuhk.edu.hk}{\nolinkurl{{hguo, luhui, wuxx, hmmeng}@se.cuhk.edu.hk}}
}

\begin{document}
%
\maketitle
\begin{abstract}
The generative adversarial network (GAN) has shown its outstanding capability in improving Non-Autoregressive TTS (NAR-TTS) by adversarially training it with an extra model that discriminates between the real and the generated speech. To maximize the benefits of GAN, it is crucial to find a powerful discriminator that can capture rich distinguishable information. In this paper, we propose a multi-scale time-frequency spectrogram discriminator to help NAR-TTS generate high-fidelity Mel-spectrograms. It treats the spectrogram as a 2D image to exploit the correlation among different components in the time-frequency domain. And a U-Net-based model structure is employed to discriminate at different scales to capture both coarse-grained and fine-grained information. We conduct subjective tests to evaluate the proposed approach. Both multi-scale and time-frequency discriminating bring significant improvement in the naturalness and fidelity. When combining the neural vocoder, it is shown more effective and concise than fine-tuning the vocoder.  Finally, we visualize the discriminating maps to compare their difference to verify the effectiveness of multi-scale discriminating.
\end{abstract}
\noindent\textbf{Index Terms}: Non-Autoregressive TTS, Speech Synthesis, Mel-Spectrogram, GAN, End-to-End Model

\section{Introduction}
\label{sec:intro}

Neural Text-to-speech (TTS) technology has achieved significant improvement with the introduction of the auto-regressive model \cite{wang2017tacotron,shen2018natural,li2019neural}. As a sequential generative model, it effectively enhances TTS in naturalness and fidelity. However, due to recursive generation and "exposure bias" \cite{guo2019new}, inference speed and stability are also affected seriously. To solve this problem, the non-autoregressive TTS (NAR-TTS) has been attracted increasing attention for better stability and parallelizability, such as \cite{FastSpeech, fastspeech2, elias2021parallel, elias21_interspeech}. It directly up-sampled the encoded text features to the frame-level sequence with the explicit duration information, then decode them to the acoustic features in parallel. But the removal of the auto-regressive mechanism also degrades generative capability, and hence affects the naturalness and fidelity. To address this problem, some generative models have been introduced, such as glow \cite{glowtts}, VAE \cite{lu21d_interspeech}, diffusion model \cite{jeong21_interspeech}, etc... Unfortunately, due to their special model design, these approaches have reduced flexibility and they cannot be easily adapted to arbitrary NAR-TTS models. 

\begin{figure}[htp]
  \centering
  \includegraphics[width=8cm]{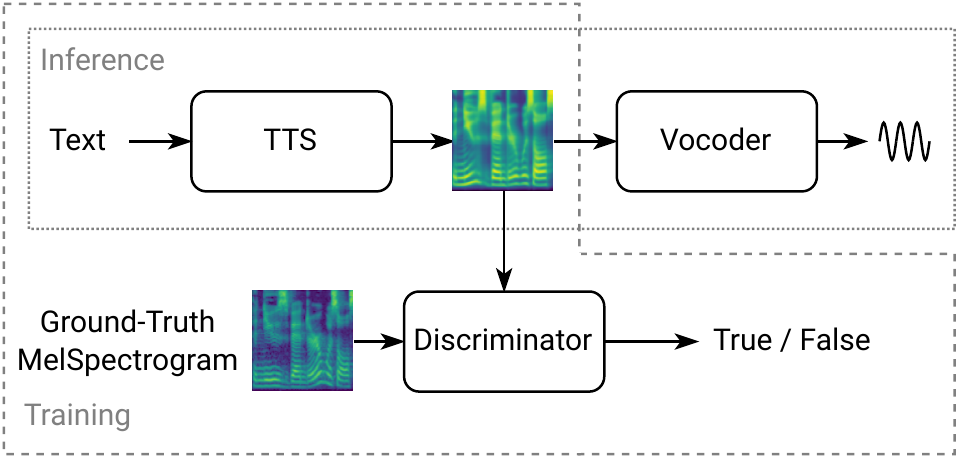}
  \caption{The framework of GAN-based TTS. The waveform denotes the generated audio. The dotted boxes indicate the modules and operations used in training or inference.}
  \label{fig:gantts}
\end{figure}

To address the aforementioned problem, the generative adversarial network (GAN) shows great potential, which has been widely applied in speech synthesis, including statistical speech synthesis \cite{8269003,saito2017statistical}, spectrogram post-filter \cite{Kaneko_2017_Interspeech}, spectrogram super-resolution \cite{MelSuperRe}, and neural vocoder \cite{kumar2019melgan, hifigantts, jang21_interspeech}. In this framework, NAR-TTS can be enhanced by only using a discriminator. As shown in Fig.\ref{fig:gantts}, in training, the discriminator is introduced to distinguish between the generated speech and authentic speech, and applied to the adversarial training to narrow the gap between these two domains. So, to maximize the benefits derived from GAN, a well-designed discriminator capturing rich distinguishable information in training is critical.

Although GAN has been widely used in neural vocoders, which adopts various structures to process the input waveform \cite{kumar2019melgan,hifigantts}, however, for NAR-TTS, the current designs of the discriminator are still preliminary, which cannot effectively incorporate the characteristics of the spectrogram. The spectrogram is a 2-D image in the time-frequency space. There is strong relationship among different elements in this image, which affects the spectral shape, harmonics, pronunciation, prosody, and timbre, locally and globally in the time-frequency space. Hence, to generate a realistic spectrogram, it is necessary to ensure that the discriminator can effectively capture this multi-scale relationship in the time-frequency domain. Different from the waveform that is down-sampled or converted to spectrograms with various STFT parameters for multi-scale discriminating, this 2-D spectrogram is hard to be processed in the same way. Alternatively, In this paper, we propose a multi-scale time-frequency spectrogram discriminator to achieve this goal. A U-Net-based structure \cite{ronneberger2015u, schonfeld2020u} is adopted to discriminate at both fine-grained and coarse-grained levels. Meanwhile, we treat the spectrogram as a 2-D image along the time and frequency axis to better exploit its information in this space.

\begin{figure*}[htp]
  \centering
  \includegraphics[width=17cm]{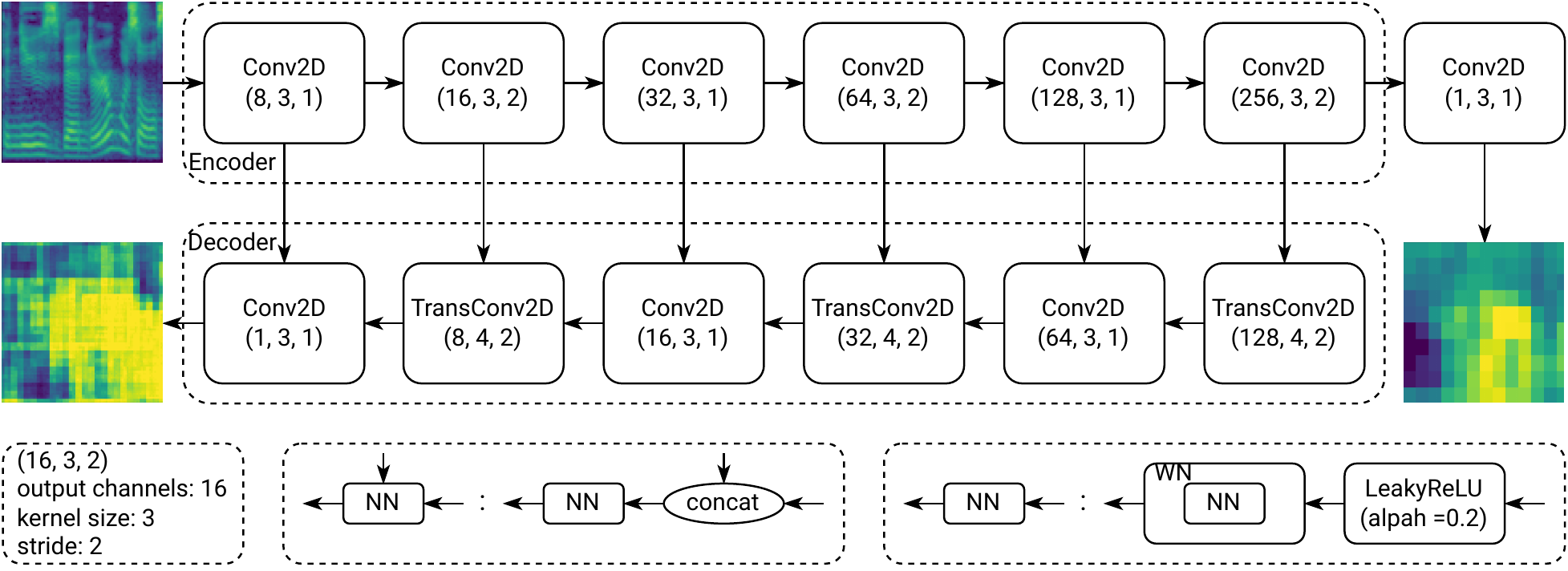}
  \caption{The architecture of the multi-scale time-frequency discriminator. The upper-left spectrogram denotes the input ground-truth or predicted Mel spectrogram. The two below heatmaps denote the fine-grained (left) and coarse-grained (right) discriminator outputs. Dotted boxes provide a detailed explanation of parameters or operations. ``concat'' denote the concatenation operation. The ``WN'' and ``NN'' denote the weight normalization layer and an arbitrary neural network layer.}
  \label{fig:disc}
\end{figure*}

This paper is organized as follows: We will first introduce our proposed discriminator, including the model structure and training algorithm for NAR-TTS. Then we will present the experiments based on ParallelTacotron, including its model structure and the specific training setup. The results of the preference tests validate that both multi-scale and time-frequency discriminating improves the training quality to NAR-TTS. When applied to the TTS system using the neural vocoder, GAN is shown as a more effective and concise approach than fine-tuning the vocoder. Finally, we visualize the output maps of the discriminator to analyze the difference between the coarse-grained and fine-grained discriminating outputs.

\section{Multi-Scale Time-Frequency Spectrogram Discriminator}
\label{sec:usd}

In this section, We will first illustrate the U-Net based model structure of the proposed discriminator, and then introduce the corresponding training algorithm to NAR-TTS.

\subsection{Model Architecture}
\label{ssec:model}

The model is illustrated in Fig.\ref{fig:disc}, which has a U-Net based encoder-decoder structure. Firstly, an encoder with a stack of convolutional layers is employed to down-sample the input spectrogram with shape $(1, T, N)$, i.e. 1 channel, T frames, N frequency bins, into the feature map with shape $(256, T/8, N/8)$. Then we use a convolutional output layer to compute the coarse-grained discriminator output map. The decoder has a structure symmetrical to the encoder, where the strided convolution is replaced with the transposed convolution. For each layer, it concatenates the output of the previous layer and the resolution-matched hidden feature map in the encoder as the input. In this way, the local high-resolution feature can be better extracted with the guide of the global information given from the encoder. After up-sampling in the decoder, a fine-grained discriminating output map with the same resolution as the input spectrogram can be achieved. Here, each convolutional layer is wrapped by a weight normalization layer \cite{salimans2016weight}, which helps stabilize adversarial training. The LeakyReLU with $\alpha=0.2$ is set as the activation function for all layers except for the input layer.

Most GAN-based vocoders are also trained with multi-scale discriminators by pre-processing the waveform into waveforms with different sample rates \cite{kumar2019melgan} or spectrograms with different STFT parameters \cite{jang2021univnet}. However, it is difficult to do so in the training of the acoustic model, since the Mel spectrogram is hard to be down-sampled well or converted to other features with different scales. The introduction of the U-Net based spectrogram discriminator makes it possible to discriminate one sequence at multiple scales directly.

\subsection{Training Algorithm}
\label{ssec:training}

We first input the text into the TTS model to generate a fake spectrogram $S_f$ for the discriminator. The target spectrogram of the text is set as the real input $S_r$. They are fed to the discriminator respectively to get the discriminating results and all hidden vectors.
\begin{equation}
    S_f = TTS(text)
\end{equation}
\begin{equation}
    C_{r}, F_{r}, H_{r} = Discriminator(S_r)
\end{equation}
\begin{equation}
    C_{f}, F_{f}, H_{f} = Discriminator(S_f)
\end{equation}
where $C_{r}, F_{r}, H_{r}$ and $C_{f}, F_{f}, H_{f}$ denote the coarse-grained output, fine-grained output, and hidden vectors of the real and fake spectrogram, respectively.

Then we update the discriminator with the LS-GAN loss function $L_D$:
\begin{equation}
\begin{aligned}
    L_d = &MSE(1, C_{r}) + MSE(1, F_{r}) \\
          &+ MSE(0, C_{f}) + MSE(0, F_{f})
\end{aligned}
\end{equation}

Before updating the TTS model, we use the updated discriminator to extract those features again, and then calculate losses as follows:
\begin{equation}
    L_{a} = MSE(1, C_{f}) + MSE(1, F_{f})
\end{equation}
\begin{equation}
    L_{f} = MAE(H_{f}, H_{r})
\end{equation}
\begin{equation}
    L_g = L_{tts} + \lambda_a L_{a} + \lambda_{f} L_{f}
\end{equation}
Adversarial loss $L_a$ is used to fool the discriminator by making $C_f$ and $F_f$ close to 1. Feature matching loss $L_{f}$ is an effective loss function to improve stablity and quality of adversrial training \cite{kumar2019melgan, yang21e_interspeech}. It calculates MAE loss for $N$ pairs of the hidden vectors of $H_f$ and $H_r$, then averages them. $L_{tts}$ is the loss function for NAR-TTS, e.g. a MSE loss between the predicted and the target spectrogram. Finally, we get $L_g$ by combining these losses with two weight parameters $\lambda_a$ and $\lambda_{f}$.

\begin{figure}[htp]
 \centering
 \includegraphics[width=0.45\textwidth]{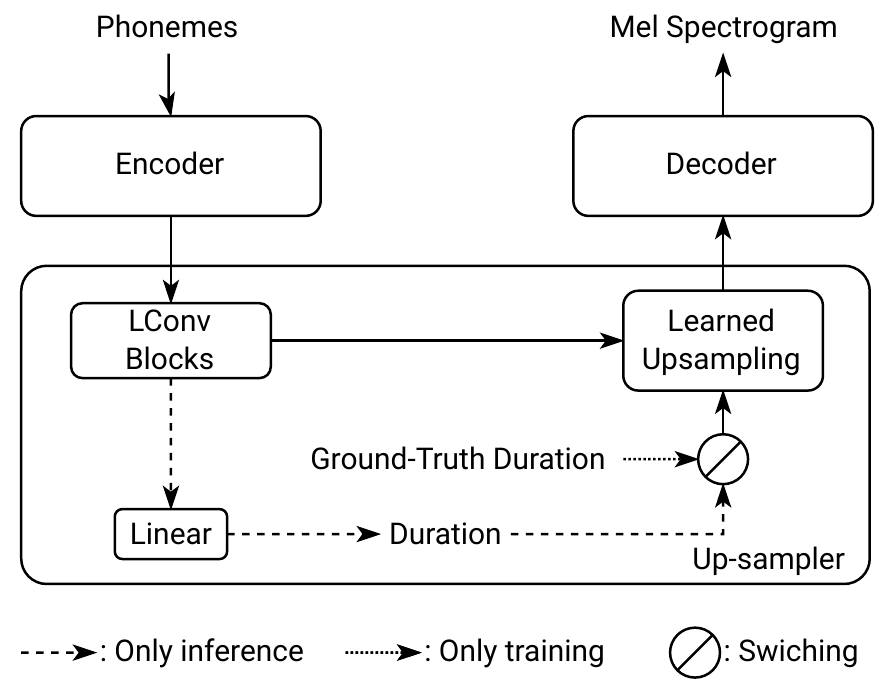}
 \caption{The structure of the simplified Parallel-Tacotron2. The phoneme sequence is processed by the encoder firstly, then further processed by LConv Blocks in the up-sampler, and up-sampled with the ground-truth or predicted durations according to the running mode (training or inference), finally decoded to the Mel spectrogram.}
 \label{fig:parataco}
\end{figure}

\section{Experimental Protocol}
\label{sec:protocol}

Our experiments are all conducted on a standard single-speaker English speech dataset, LJSpeech, with over 10 hours of recordings. After screening and pre-processing, we collect 11000 pairs of (text, 16kHz audio) as the training set.

\subsection{Non-Autoregressive TTS}
\label{ssec:tts_model}

As shown in Fig.\ref{fig:parataco}, the model is implemented based on ParallelTacotron2 \cite{elias21_interspeech}, but removes speaker embedding and residual encoder for simplification. In this model, the learned upsampling module can up-sample the input, a phoneme sequence with punctuations, to the frame-level features according to explicitly predicted phoneme-level durations. Then we use the decoder to generate the 80-dim log-scale Mel-spectrogram with 12.5ms frameshift and 50ms frame length.

Notably, we avoid using Soft-DTW loss in our experiments due to its huge cost on computing resources. Instead, in the training stage, we use ground-truth duration\footnote{We get the duration using MFA at \url{https://github.com/MontrealCorpusTools/Montreal-Forced-Aligner}} as input and add an extra loss function between the predicted and target duration. It is also an effective approach with less computing cost for duration learning. To reconstruct waveform from Mel-spectrogram, Griffin-Lim \cite{griffin1984signal} and Hifi-GAN \cite{hifigantts} trained on the same dataset are used in the tests\footnote{The code of HifiGAN is available at \url{https://github.com/jik876/hifi-gan}}.

\subsection{Training Setup}
\label{ssec:setup}

We use MSE and MAE loss functions for the iterative loss of Mel-spectrogram and the duration loss as follows:
\begin{equation}
    L_{spec} = MSE(S, \hat{S}) + MAE(S, \hat{S})
\end{equation}
\begin{equation}
    L_{dur}  = MSE(D, \hat{D}) + MAE(D, \hat{D})
\end{equation}
\begin{equation}
    L_{tts}  =  L_{spec} + \lambda_{dur} L_{dur}
\end{equation}
where $L_{spec}$ denotes the loss function between the target spectrogram $S$ and the predicted spectrograms $\hat{S}$, $L_{dur}$ denotes the loss function between the target duration $D$ and the predicted duration $\hat{D}$. $\lambda_{dur}$ is set to balance these two losses, which is 0.02 in our experiments. The output Mel-spectrogram of the last layer in the decoder is involved in the adversarial training. $\lambda_a$ and $\lambda_f$ are set to 0.2 and 2 for all experiments with U-Net based discriminators, otherwise are 1 and 10.

RAdam \cite{liu2019variance} with $(\beta_1=0.9, \beta_2=0.999)$ and Lookahead \cite{lookaheadopt} with $(k=5, \alpha=0.5)$ are combined as the optimizer to provide more stable training process. The learning rate is exponentially decayed from $1e^{-3}$ to $1e^{-5}$ after 20,000 iterations. All models are trained for 200,000 iterations with a batch size of 64. 

\section{Results}
\label{sec:results}

We conduct subjective tests using Amazon Mechanical Turk. 80 utterances in the dataset which are disjoint from training set are used as the test set. For each test, each listener can only rate 30 sets of utterances to ensure good test quality. \footnote{Samples are available at \url{https://hhguo.github.io/DemoUGANTTS}}

\begin{figure}[htp]
  \begin{subfigure}[htp]{0.45\textwidth}
     \centering
     \includegraphics[width=1.0\textwidth]{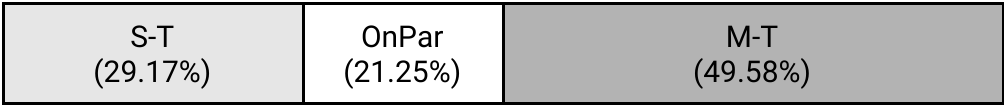}
     \caption{The comparison between S-T and M-T ($p$-value=$1.58e-4$) \\ \quad}
     \label{fig:ms}
  \end{subfigure}
  \begin{subfigure}[htp]{0.45\textwidth}
     \centering
     \includegraphics[width=1.0\textwidth]{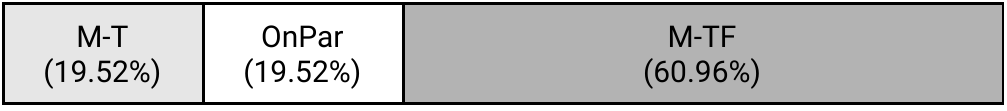}
     \caption{The comparison between M-T and M-TF ($p$-value=$7.63e-13$)}
     \label{fig:tf}
  \end{subfigure}
  \caption{The preference test for different discriminators}
  \label{fig:cmos}
\end{figure}

\subsection{Discriminators}
\label{ssec:disc_mos}

The preference tests are conducted to validate that multi-scale and time-frequency discrimination are both effective for the adversarial training. Three discriminators are involved in the comparison:
\begin{enumerate}
    \item Single-Scale Time Discriminator (S-T), which only uses the encoder part and 1-D convolutions along the time axis. It has been validated effective in \cite{yang21e_interspeech}.
    \item Multi-Scale Time Discriminator (M-T), which is based on S-T, and uses both encoder and decoder.
    \item Multi-Scale Time-Frequency Discriminator (M-TF)
\end{enumerate}
We first compare S-T and M-T, then compare M-T and M-TF. Here, all samples are generated using Griffin-Lim \cite{griffin1984signal} to avoid the bias brought from data-driven neural vocoder.

In the comparison between S-T and M-T shown in Fig.\ref{fig:ms}, M-T achieves significant preference with the voting rate of 49.58\%. We find that they generate similar timbre and rhythm overall, but multi-scale discrimination makes the formant smoother and clearer, and produces better prosody with higher naturalness and diversity, hence receives higher listener preference. The comparison between M-T and M-TF in Fig.\ref{fig:tf} validates that the time-frequency operation is an effective approach by the higher voting rate of 61.11\%. It obviously improves the fidelity, including the spectral clarity, smoothness, and continuity, which can be easily noticed by listeners. It shows that operating the spectrogram as a 2-D image along both time and frequency axes can better exploit spectral information.

\begin{table}[htp]
\centering
\caption{The MOS test results (\checkmark and $\times$ denote the corresponding approach is used or not. $\pm$ indicates 95\% CI)}
\begin{tabular}{ccc}
\textbf{GAN} & \textbf{Finetune} & \textbf{MOS} \\ \hline
$\times$    & $\times$     & $2.91 \pm 0.14$ \\ 
$\times$    & \checkmark   & $3.43 \pm 0.15$ \\ 
\checkmark  & $\times$     & $\mathbf{3.81 \pm 0.15}$ \\ 
\checkmark  & \checkmark   & $3.71 \pm 0.14$ \\ \hline
\multicolumn{2}{c}{Analysis-Synthesis} & $3.97 \pm 0.17$ \\ \hline
\end{tabular}
\label{tab:mos}
\end{table}

\subsection{GAN or Fine-tuning Vocoder? Or Both?}
\label{ssec:finetune}

In mainstream TTS systems, the neural vocoder is usually adopted for waveform generation due to its high-quality generation. It is trained with ground-truth spectrograms but is used based on the predicted ones. The gap between them causes errors in vocoding, such as noise and distortion. To narrow the gap, fine-tuning the vocoder with TTS predictions is often used, but also leads to more costs on computing and storage resources, and more complicated TTS training and update. Instead, it is more concise to directly improve the fidelity of the generated spectrogram, e.g. GAN training. In this paper, both our approach and fine-tuning vocoder are evaluated in this aspect via a MOS test.

\begin{figure}[htp]
  \centering
  \includegraphics[width=8cm]{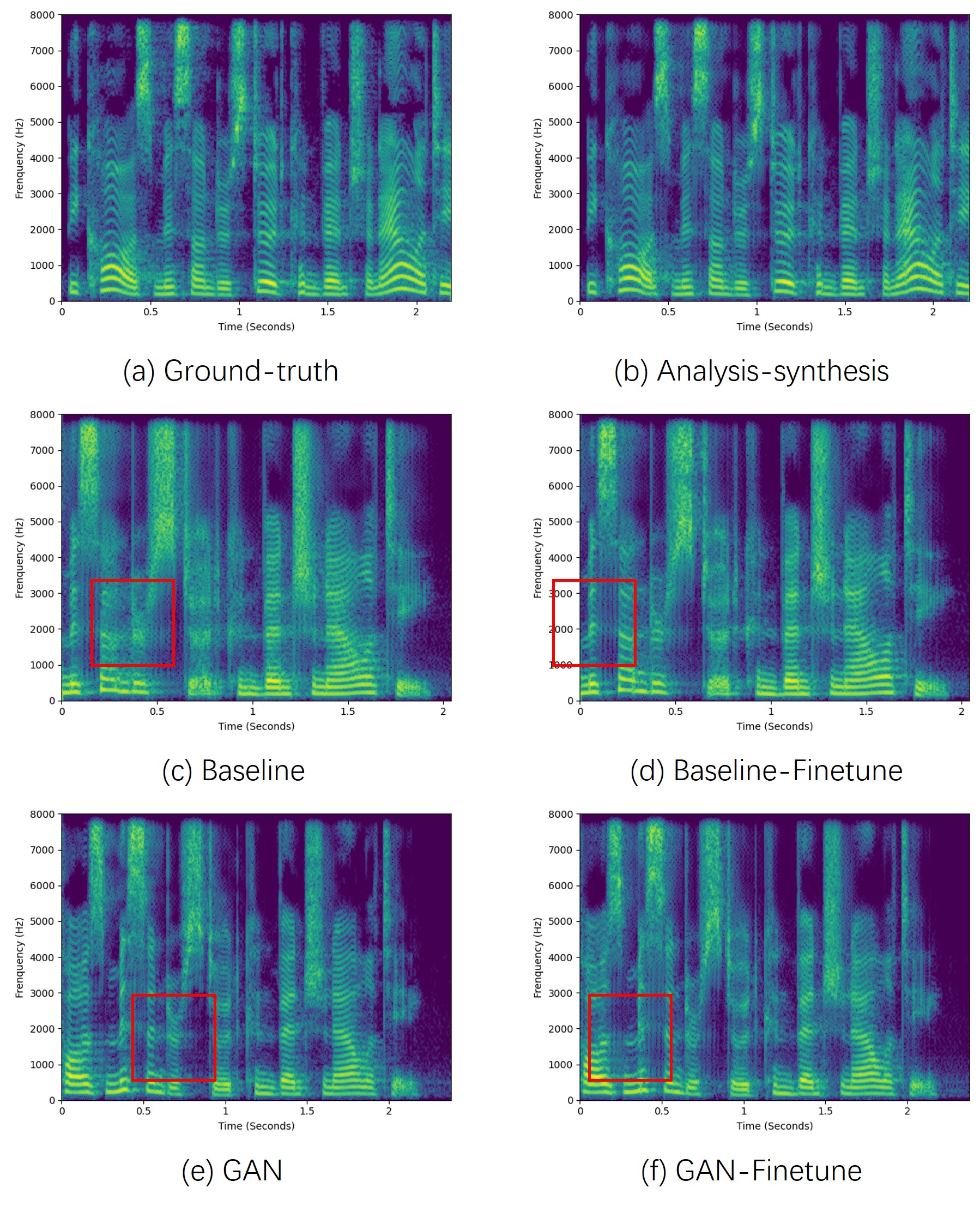}
  \caption{The magnitude spectrograms of the ground-truth audio (a), the analysis-synthesis audio (b) and audios generated by baseline models w/ (c) or w/o (d) fine-tuned vocoder (d), GAN-based models w/ (e) or w/o (f) fine-tuned vocoder.}
  \label{fig:spec}
\end{figure}

The results of the MOS test and spectrogram generated by these models are shown in Table.\ref{tab:mos} and Fig.\ref{fig:spec}, respectively. Firstly, the high-fidelity reconstruction of analysis-synthesis in Fig.\ref{fig:spec}(b) shows that the vocoder is trained well. The baseline system, without GAN training and fine-tuned vocoder, receives the worst score of 2.91. The fuzzy spectrogram with unsmoothed low-frequency harmonics leads to serious degradation in naturalness and fidelity, which is shown in Fig.\ref{fig:spec}(c). After fine-tuning, the harmonics in the middle and low-frequency parts are enhanced significantly, hence improving its output quality with a much higher MOS of 3.43.

For the GAN-based system, it can already achieve a much higher score of 3.81 without fine-tuning. The spectrogram shown in Fig.\ref{fig:spec}(e) presents both clearer, smoother harmonics and richer information in high frequency. However, the system using both GAN and fine-tuning achieves a worse score of 3.71. The harmonics in the middle-frequency part become slightly fuzzier, hence degrading the fidelity. We find that, when the vocoder is fine-tuned to adapt predicted spectrograms, the larger the discrepancy between the ground-truth spectrogram and the predicted one, the fine-tuning is harder. Although GAN training reduces the distribution gap between these two domains, the distance between these two spectrograms is also enlarged. It may degrade the fidelity of the vocoder. In conclusion, GAN can bring benefits to synthesis, while finetuning may not bring consistent.

\subsection{Discriminating Visualization}
\label{ssec:analysis}

\begin{figure}[htp]
  \centering
  \includegraphics[width=8cm]{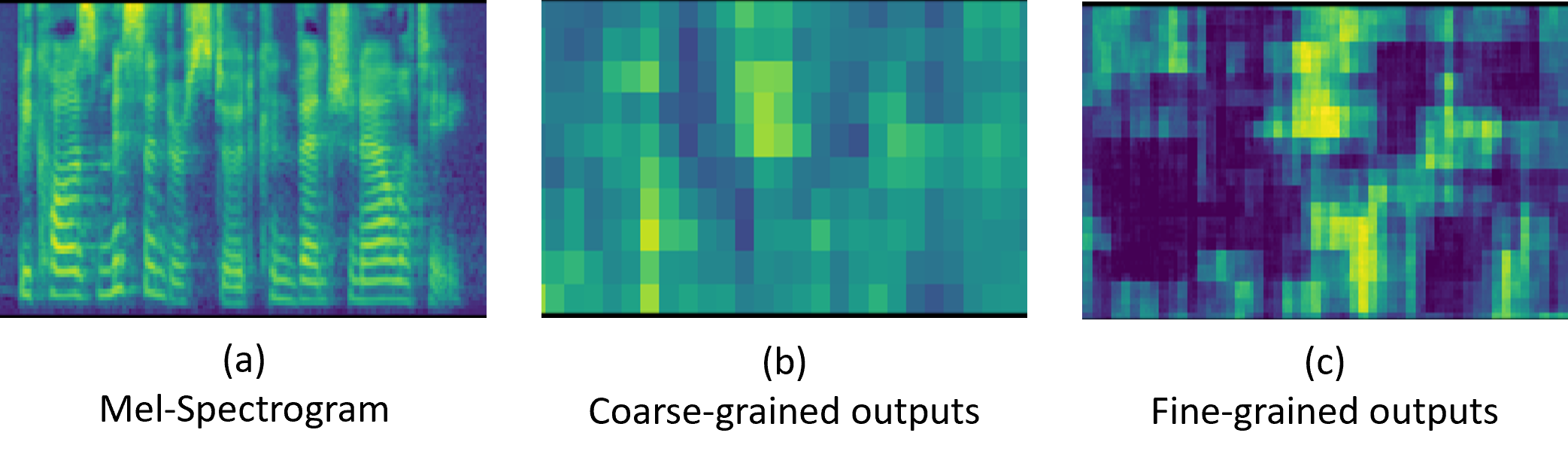}
  \caption{The visualization of the multi-scale time-frequency discriminator outputs.}
  \label{fig:vis}
\end{figure}

As shown in Fig.\ref{fig:vis}, we visualize two output maps of the multi-scale time-frequency discriminator. The high-lightness area indicates that it has a higher probability classified as the real one. The up-sampled coarse-grained output map in (b) shows a smooth and averaged heatmap, which provides coarse-grained, global discriminating information. In comparison, (c) shows a sharper heatmap that has higher resolution and more attention on local parts. This fine-grained discriminating enhances fidelity-related information like formants.

\section{Conclusion}
\label{sec:conclusion}

This paper proposes a multi-scale time-frequency spectrogram discriminator to provide better GAN training for Non-Autoregressive TTS. It operates the Mel-spectrogram in time-frequency domain at different scales to exploit richer information for better discrimination. The preference tests validate the effectiveness of multi-scale and time-frequency discriminating. An MOS test is conducted to investigate the impact of GAN and vocoder fine-tuning on NAR-TTS. The results show that GAN training significantly improves TTS with the higher MOS of 3.81, but fine-tuning may have negative effects. In addition, the coarse-grained and fine-grained discriminator output maps are visualized to investigate their differences, and verifies that they can provide richer discriminative information at different scales.

\vfill\pagebreak

\bibliographystyle{IEEEtran}
\bibliography{refs}

\end{document}